\documentclass[conference]{IEEEtran}
\IEEEoverridecommandlockouts
\usepackage{cite}
\usepackage{amsmath,amssymb,amsfonts}
\usepackage{algorithmic}
\usepackage{graphicx}
\usepackage{textcomp}
\usepackage{xcolor}
\usepackage[normalem]{ulem}
\usepackage{url}
\usepackage{tikz, enumitem}
\usetikzlibrary{positioning}
\usepackage[]{algorithm2e}

\def\BibTeX{{\rm B\kern-.05em{\sc i\kern-.025em b}\kern-.08em
    T\kern-.1667em\lower.7ex\hbox{E}\kern-.125emX}}
\begin{document}

\title{{\it Cross Hashing}: Anonymizing encounters in Decentralised Contact Tracing Protocols}

\author{\IEEEauthorblockN{Junade Ali}
\IEEEauthorblockA{\textit{University of Bedfordshire}\\
Luton, United Kingdom \\
junade.ali@study.beds.ac.uk}
\and
\IEEEauthorblockN{Vladimir Dyo}
\IEEEauthorblockA{\textit{University of Bedfordshire}\\
Luton, United Kingdom \\
vladimir.dyo@beds.ac.uk}
}

\maketitle

\begin{abstract}
During the COVID-19 (SARS-CoV-2) epidemic, Contact Tracing emerged as an essential tool for managing the epidemic. App-based solutions have emerged for Contact Tracing, including a protocol designed by Apple and Google (influenced by an open-source protocol known as DP3T). This protocol contains two well-documented de-anonymisation attacks. Firstly that when someone is marked as having tested positive and their keys are made public, they can be tracked over a large geographic area for 24 hours at a time. Secondly, whilst the app requires a minimum exposure duration to register a contact, there is no cryptographic guarantee for this property. This means an adversary can scan Bluetooth networks and retrospectively find who is infected. We propose a novel "cross hashing" approach to cryptographically guarantee minimum exposure durations. We further mitigate the 24-hour data exposure of infected individuals and reduce computational time for identifying if a user has been exposed using $k$-Anonymous buckets of hashes and Private Set Intersection. We empirically demonstrate that this modified protocol can offer like-for-like efficacy to the existing protocol.
\end{abstract}


\section{Introduction}

Contact tracing systems are based on sharing of individual's presence through short-range Bluetooth beacons and enable individuals to determine if and when they have been in contact with an infected person \cite{cho2020contact}. 
If used by enough people, epidemic control can be achieved without resorting to country-wide "lockdown" strategies which are harmful to society and the economy \cite{ferretti2020quantifying}. 
At the same time pervasive tracking of people locations raised serious privacy concerns globally including potential risk to {\it human rights}  with claims that excessively compromising privacy is a gateway to undermining other rights  \cite{hrw}. 

While a number of privacy-aware protocols have been proposed, many do not protect the privacy of the person once they becomes infected and allow sharing of their location information. 
The area is rapidly evolving with numerous privacy aware protocols developed and deployed up to date \cite{vaudenay2020analysis} \cite{applegoogle}. 
In particular, Apple and Google, who control the overwhelming market share of mobile phone operating systems, have created a BLE-based protocol for contact tracing [5], based on DP3T, an open-source contact tracing protocol using hash-based cryptography to achieve anonymity. 
The protocols protect the users' privacy by encrypting their beacon IDs with temporal daily keys, which are created by the individual and can be disseminated in the event a user tested positive. 
However, once an infected user shares his daily temporal key, it becomes possible to track his or her entire  daily trajectory. 
In fact, a formal analysis of DP3T  has revealed that  it is possible for "malicious users to re-identify infected users at scale" \cite{vaudenay2020analysis}.
As the locations of infected individuals tend to attract extensive news coverage and re-identification leading to unwanted privacy invasion and public distain, this represents a serious issue \cite{lewer2019clinic}.  

Another concern is that the location information of an infected person becomes available to anyone who recorded at least one beacon, i.e. passersbys or Bluetooth trackers installed in various locations, who have not been in a meaningful contact with the infected person and therefore are not supposed to have access to the infected person's information. 

In this paper, we propose a privacy aware contact tracing approach that ensures the privacy of the person even after they become infected. 
The novelty of the approach is that we choose to protect the proximity information not at the individual beacon level but on the {\em encounter} level of a certain biologically meaningful duration. 
The approach ensures that location information is shared only with individuals who have spent a minimum duration in the proximity of the infected individual, reducing the risk of  re-identification attacks.  
Furthermore, by cryptographically signing encounters the approach ensures that once the infected individual shares his or her location information, it is not possible for a malicious user to re-identify the person or reconstruct his trajectory. 



\section{Related Work}

Re-identification attacks of anonymised datasets have been explored before in literature, a description of the problem can be found in \cite{ohm2009broken}. During the 2020 COVID-19 outbreak, \cite{park2020information} notes a number of privacy controversies in South Korea through the implementation of Information Technology contact tracing systems; with the locations of infected individuals attracting extensive news coverage and re-idenitification allegedly leading to unwanted privacy invasion and public distain.

The \cite{applegoogle} protocol by Apple and Google works by generating a Temporary Exposure Keys (otherwise known as a Daily Tracing Key, a random number which is SHA-256 hashed and truncated to 16 bytes), this daily key is then in turn used to generate Rolling Proximity Identifiers for each 10 minute interval in the 24 hour period. The Rolling Proximity Identifiers is created by concatenating the Temporary Exposure Key with a timestamp value, computing a SHA-256 hash and truncating it to 16 bytes. In the event an infection is detected, the Temporary Exposure Keys are distributed and users can generate each Rolling Proximity Identifiers to check if they encountered an infected individual.

The \cite{applegoogle} protocol shares the same flaw as DP3T in that the Temporary Exposure Keys are widely circulated, allowing for re-identification of infected users. Finally, the protocol derives new Rolling Proximity Identifiers at a fixed time interval regardless of other characteristics (such as distance travelled), raising the possibility of tracing journeys travelled by users.

The key issue to solve here is allowing for users to compute the Rolling Proximity Identifiers of users they may have come into contact with without leaking the information of all affected users. Developers of contact tracing apps \cite{ROBERT} have sought guidance on implementing our original $k$-Anonymity communication protocol \cite{ali2017mechanism} to mitigate this effect (a protocol originally devised for compromised credential checking).

It is noted in \cite{gvilisecurity} that this risk is not sufficiently mitigated in existing protocols; the paper discusses that an attacker can passively collect proximity identifiers and then target that person if their diagnosis key is later disclosed and note that $k$-Anonymity may offer a solution to this problem. This risk is further discussed in \cite{vaudenay2020analysis} and notes that this is a substantial problem to address, noting "the only way to mitigate this attack is to deny the same privileges as the apps to individuals" and suggesting it could require a challenging deployment of hardware encryption (TPM chips) to address.

Our prior work in \cite{li2019protocols} provides an empirical comparison of {\it compromised credential checking} (C3) protocols and defines novel protocols for minimising information loss. Whilst pure Private Set Intersection has a heavy computational and communication overhead, \cite{thomas2019protecting} has combined $k$-Anonymous protocols with Private Set Intersection to reduce this burden. Most recently support for response padding has been introduced into the original $k$-Anonymity protocol to prevent inference on the wire of encrypted contents \cite{pwnedpasswordspadding}.

Wireless scanner equipment using MAC address tracking has been deployed for monitoring pedestrian and cyclist journey time, monitoring road conditions in real-time and passenger movements throughout train stations. Such continuous tracking over a large geographical scale has raised serious privacy concerns amongst governments and the general public. In \cite{ali2020practical}, we provided a novel practical hash-based approach to anonymising such MAC Addresses at the point of collection. This approach was designed for use-cases like Journey Time Monitoring Systems which do not rely on client-side software to allow for the anonymisation of data, nevertheless the paper provides a novel approach for anonymising wireless unique identifiers, which is of relevance to this work.

\section{Threat Model}

The Contact Tracing protocol described in \cite{applegoogle} works by a user randomly generating a Temporary Exposure Keys (formerly known as a Daily Tracing Key) which is 16 bytes long and generated from a cryptographic random number generator (CRNG). The device stores one Temporary Exposure Key per day up to a maximum of 14 keys (covering a 14 day period).

A Rolling Proximity Identifier (RPI) is derived by hashing the Temporary Exposure Key together with an integer count of the 10 minute interval associated to that 14 day period. This RPI is broadcast and collected by other users.

In the event the user is found to have tested positive for the virus, they may disclose their Temporary Exposure Keys to a server. Users may then download a list of Temporary Exposure Keys and compute the RPIs for each 10 minute interval to see if they were exposed to the virus.

An adversary may passively scan for RPIs broadcast over a geographic area. Using lists of Temporary Exposure Keys they are able to identify and trace infected individuals throughout a large geographic area for that 24 hour period using sensors installed on public transport networks, motorway Bluetooth Journey Time Monitoring Systems or even rubbish bins \cite{ali2020practical}.

Accordingly, the threat is not merely that an adversary may identify an infected individual, but that they may also trace them over a large geographic scale for a 24 hour period.

The \cite{applegoogle} protocol requires public health authorities set a minimum threshold for time spent together for an encounter to the counted, this value must be greater than 5 minutes. It is important to note that an adversary can passively collect RPIs without any need to have due regard for the minimum exposure threshold.


\section{Anonymity Model} \label{anonymity-model}

We seek to limit information leakage by the server containing the Daily Tracing Keys of users whilst allowing users to validate if they have been exposed. We propose two modifications to the protocol to enhance privacy; the first provides a cryptographic guarantee that a user must be contact with another for a minimum duration before a contact is established and the second provides efficient lookups of RPIs without the need to disclose the Temporary Exposure Keys.

\subsection{Cryptographic Contact Duration Threshold}

In the current protocol, contact duration thresholds are set through software configuration rather than part of the cryptographic specification. By providing a cryptographically guaranteed contact duration threshold, we are able to limit information exposure before that threshold is met.

We propose that the Rolling Proximity Identifiers (RPIs) are rotated at the same frequency as the minimum contact threshold. Contact events are registered by computing a hash of the current RPI and one of the previous RPI value, we refer to this as a Consistent Contact Identifier (CCI). To identify exposures later, these CCIs are searched from a remote server.

Where RPIs generated from a device during the course of a day are represented as $RPIs = \{i_1, i_2,..., i_n\}$, a CCI for a given period can be computed as $CCI = HKDF(i_n, i_{n-k}), k \geq 1$ where $k$ is the number of RPI steps required to achieve the minimum contact threshold and $HKDF$ refers to the HKDF hash function used by the \cite{applegoogle} protocol (adopted from IETF RFC 5869). Where multiple steps are required, $k$ may be defined as a set of values.

In lieu of a user downloading a list of Temporary Exposure Keys from users who are exposed to the virus, we propose users instead obtain a list of CCIs of exposed users. This prevents a user being identified as exposed unless the interaction was for the minimum contact duration threshold and constrains any tracking to the period of time around the immediate contact. By {\it cross hashing} RPIs together, we achieve a minimum time guarantee using standard cryptographic primitives.

If only the RPIs are broadcast (as is the case in the current protocol) and there are no changes made to the Bluetooth protocol - the number of possible CCIs increases exponentially. To mitigate this performance disadvantage, we propose a simple change to the beacon broadcast by the devices. The devices currently broadcast the 16 byte RPI key; instead we broadcast a $12$ byte RPI key, allowing up to two lots of $2$ byte CCI prefixes to also be broadcast acting as a signature. These signatures allow the device to rapidly compute which RPIs produce valid CCIs and discard nonsensical values. The token that is broadcast is shown in Fig \ref{fig:broadcast}. Such signatures are simply generated by the device computing the $CCI$ value for one and two intervals in the past and truncating the value to $2$ bytes.

Algorithm \ref{alg:validate} shows how this approach can be utilised to identify which CCIs should be stored as valid values, and looked for in the event they become positive later. As the RPI and signature pair can be represented in $14$ bytes, there is a minor decrease in entropy from the old $16$ byte RPI value, but should should not cause any issues in ordinary usage \cite{ali2020practical}. As the $2$ byte signature is capable of storing $65,536$ distinct values, there is a small risk of a collision, but this would merely result in a small amount of excess nonsensical data being stored on a users device.

\begin{figure}
\centering
\begin{tikzpicture}
  [
    my text node/.style={draw, very thick, align=center}
  ]
  \node (example-align) [my text node, text width=10em] {RPI \\ 12 bytes};
  \node (example-align2) [my text node, right=0cm of example-align.north east, anchor=north west, text width=5em] {CCI Prefix$_1$ \\ 2 bytes};
  \node (example-align3) [my text node, right=0cm of example-align2.north east, anchor=north west, text width=5em] {CCI Prefix$_2$ \\ 2 bytes};
\end{tikzpicture}
\caption{BLE broadcast of $16$ bytes, consisting of the devices RPI associated with that timestamp (of 12 bytes) and up to $2 \times 2$ byte CCI prefixes.}
\label{fig:broadcast}
\end{figure}

\begin{algorithm}[]
 \KwData{Historic RPIs, Current RPIs}
 \KwResult{Store CCIs = []}
 initialization\;
 \For{RPI in Current RPIs}{
  \For{Last RPI in Historic RPIs}{
    CCI = HKDF(truncate(RPI, 0, 12), truncate(Last RPI, 0, 12)) \\
    signatures = explode(truncate(Last RPI, 12, 16), 2)
    \If{truncate(RPI, 0, 2) in signatures}{
    	Store CCIs += CCI
    }
  }
 }
 \caption{This algorithm demonstrates how from an input of $Historic RPIs$ and an input of $Current RPIs$, it is possible to produce an array of CCIs, $Store CCIs$, which can then be searched later in the event of a positive test.}
 \label{alg:validate}
\end{algorithm}

\subsection{Efficient Rolling Proximity Identifier Search}

The advantage of supplying and downloading the daily Temporary Exposure Keys to users is that there is a lower communication overhead than downloading all RPIs or our proposed CCIs. With RPIs rotated every 10 minutes, there are 144 times the number of keys to download; and if rotated every 5 minutes, this increases to 288 times the original. In our approach; as the number of CCIs grows exponentially to the number of neighbours (if a device is not able to identify which neighbours rotated their IDs at a particular time), efficient search is increasingly important.


We seek to allow a user to query CCIs whilst minimising the privacy loss both on the server and for the client. Fortunately, this is a well explored problem in \cite{li2019protocols, ali2017mechanism, thomas2019protecting}.

When a user's Temporary Exposure Key is uploaded to a server, the CCIs are computed for the period of exposure. They are then stored in $k$-Anonymous buckets described in \cite{ali2017mechanism}. The client simply needs to truncate the CCI hash to a desired number of bits such to grant anonymity (we provide both formal definitions and empirical analysis for applying $k$-Anonymity to hash based data sets in \cite{ali2020practical}). As such buckets are queried on the basis of CCI hashes instead of the RPIs or Temporary Exposure Keys, no additional information is leaked (for more information on data leakage risks see the section on "Identifier-Based Bucketization" in \cite{li2019protocols}).

As a further safeguard against excessive data leakages, look-ups within those buckets are performed using Private Set Intersection (PSI). An approach for implementing Private Set Intersection in $k$-Anonymous buckets is described in further detail in \cite{thomas2019protecting} using a Diffie-Hellman based approach. Private Set Intersection allows two parties to determine the intersection of two sets of data without disclosing either set to the other party. During a $k$-Anonymous search protocol, this means the client discloses a partial hash as the search partition and then both parties are able to determine if the content of the clients search is in that partition without needing to disclosure any further information. Unlike the approach detailed in \cite{ali2017mechanism}, this provides further cryptographic shielding of the hashes held by the server.

As a final security measure, to prevent an adversary determining the number of daily contacts through passive analysis (as analysed in\cite{pwnedpasswordspadding}), provision should be made for the number of requests to be padded with random data.

\section{Evaluation}

We seek to evaluate and quantify any decrease in the number of contacts registered through the use of our {\it cross-hashing} approach. For evaluation, we use the BLE Beacon dataset \cite{sikeridis2018blebeacon} which consists of real-world traces from users carrying BLE beacons within a multi-floor data university building. It is important to note that this dataset is based on beacons collected by fixed-position sensors from mobile devices carried by users, instead of being both broadcast and collected from mobile devices. We implement upon this dataset both the unmodified protocol and the amended version detailed in Section \ref{anonymity-model}.


The dataset consists of a total of $18673095$ beacon detection events across $36$ days, using a periodic transmission rate of 1 Hz (with power of 0 dBm). Given the current protocol defines that scanning should occur once at least every 5 minutes, with a 10 minute minimum exposure threshold \cite{applegoogleexposure}, we consider a device as lost when more than 15 minutes have elapsed between consecutive beacons (3 scanning intervals). This amounted to a total of $16355$ daily interactions between unique devices. Of these, $7800$ ($48\%$) contained events which met our minimum contact threshold requirement of at least 2 beacons being detected at least 10 minutes apart. As part of this experiment, we do not "down sample" the original dataset as the RPI rotation time is fully independent from the actual beacon rate in the protocol.

Recall from Section \ref{anonymity-model} that a Consistent Contact Identifier (CCI) is computed using the number of steps, $k$, of Rolling Proximity Identifiers (RPIs) required to achieve the minimum contact threshold. We evaluate the coverage obtained by $k = 1$ (the last 5 minute interval), $k = 2$ (the 5 minute interval before last) and $k = \{1, 2\}$ (either in the last 5 minute interval or the one prior to that).

\begin{table}[b]
\centering
\caption{A $95\%$ capture rate is achieved by $k = 1$ and $100\%$ by $k = \{1, 2\}$.}
\label{tab:results}
\begin{tabular}{lll}
RPI Steps ($k$) & Events Captured & Capture Rate \\
$k = 1$         & $7430$            & $95\%$         \\
$k = 2$         & $6027$            & $77\%$         \\
$k = \{1, 2\}$  	& $7800$            & $100\%$       
\end{tabular}
\end{table}

The results of the evaluation are summarised in Table \ref{tab:results}. A single RPI step ($k = 1$) provides 95\% coverage of daily contact events. A configuration of $k = \{1, 2\}$ (whereby two steps are computed) achieves $100\%$ coverage. Obtaining full coverage by using the $k = \{1, 2\}$ configuration instead of $k = 1$ provides the missing $5\%$ coverage, but requires twice the storage and computational overhead.

As the original dataset captures beacons without any range restriction, the $5\%$ missing interactions where $k = 1$ is likely the result of percentage of beacons coming in and out of contact. This is less likely in close encounters but could have epidemiological impact and therefore may be beneficial for the original specification to clarify the impact of lost beacons on the minimum contact threshold.

\section{Down Sampling Simulation}

So far we have evaluated the impact of cross-hashing on a dataset collected using a periodic transmission rate of 1 Hz (with power of 0 dBm). For mobile devices, even using Bluetooth Low Energy, it may not be energy efficient to broadcast beacons constantly. Manufacturers of mobile devices may instead wish to time the broadcast of beacons with the sleep/wake behaviour of the device, subject to a minimum in a given time interval.

We re-run the previous experiment without any impact to the transmission rate or power, but only allowing the beacon to the broadcast for a limited duration of time which is randomly determined within the minimum contact threshold.

For broadcast times of $30$ seconds, $1$ minute, $2$ minutes and $5$ minutes (in every $10$ minutes); we evaluate how the total events and capture rate against the existing protocol, as well as evaluating the percentage of events lost overall by downsampling. The results are shown in Table \ref{tab:downsampling}.

\begin{table}[]
\centering
\caption{Our cross-hashing approach remains equally effective even when the beacon broadcast time is less than the minimum contact threshold.}
\label{tab:downsampling}
\begin{tabular}{llll}
Beacon Broadcast Time & Valid Events & Capture Rate & Events Lost \\
$0.5$ mins   & $365$     	& $100\%$	& $95\%$       \\
$1$ mins     & $733$     	& $100\%$	& $91\%$       \\
$2$ mins  	 & $2283$     	& $100\%$	& $71\%$       \\
$5$ mins   	 & $6234$     	& $100\%$	& $20\%$
\end{tabular}
\end{table}

In all instances, our cross-hashing configuration of $k = \{1, 2\}$ was still able to perform $100\%$ as effectively as the existing approach, but down sampling the broadcast rate had a substantial impact on the percentage of beacons lost. When the beacon is only broadcast for $5$ out of every $10$ minutes, only $20\%$ of events are lost but $71\%$ of events are lost when this is reduced to $2$ in every $10$ minutes. Nevertheless, this phenomenon is out of scope for this work but may be interesting to study in the future.

This section does however demonstrate that, despite downsampling the beacon broadcast time within the minimum contact threshold, our cross-hashing approach works achieves like-for-like coverage when compared with the existing approach.

\section{Security Discussion}

Our proposed amendments to the protocol mitigates passive de-anonymisation attacks in two ways; the first is by requiring a contact to last for longer than the minimum duration for a user to know if another user is infected and the second prevents whole-day tracking in instances where the user is infected.

The {\it cross hashing} approach used to mitigate such attacks necessitates more hashes per infected user (protecting information at an {\it encounter} level) and accordingly a higher communication overhead. We mitigate this by signing the BLE broadcast tokens with $2$ byte prefixes of valid CCIs. We then allow the client to search in $k$-Anonymous buckets which reduces the communication overhead and reduces data leakage by the server through the use of Private Set Intersection. Nevertheless, this results in some data leakage.

There is a trade-off to this approach; to ensure both communication overhead remains low and prevent re-identification attacks, the user must disclose the first few bits of two hashed Rolling Proximity Identifiers. The risk associated to such a protocol is well documented in \cite{li2019protocols} and formal analysis of the properties of anonymised hashes can be found in \cite{ali2020practical}. Such protocols have been used widely in compromised credential checking systems, a privacy sensitive context.

Nevertheless, even without preventing such de-anonymisation attacks; communication overhead may be a future problem in the existing protocol as it grows linearly to the number of positive tests (regardless of user interactions). Once a user is infected, a Temporary Exposure Key of 16 bytes must be stored for 14 days. This means the communication overhead of downloading Temporary Exposure Keys exceeds 100 MB after $446429$ users have tested positive. Similarly, depending on the number of infected users and communication limitations, {\it cross hashing} may be deployed as a standalone measure.

This approach mitigates, but does not eliminate, de-anonymisation attacks. An active attacker who approaches a smartphone for longer than the Contact Duration Threshold can still determine if the user later tested positive. We do however eliminate the risk of tracking over a 24 hour period.

\section{Conclusion}

In this paper we have provided a practical approach for mitigating passive de-anonymisation attacks on the Google/Apple Contact Tracing Protocol \cite{applegoogle}. We provide a novel {\it cross hashing} approach to provide a cryptographic guarantee that a contact is only registered when two devices have been in contact for a minimum duration. We further reduce the communication overhead of this protocol using $k$-Anonymity buckets of hashes and Private Set Intersection.

The work in this paper provides a tangible proposal rectifying weaknesses in the anonymity scheme used in the existing protocol. Real-world BLE Beacon traces indicate this approach can enable like-for-like detection efficacy of the existing protocol.

There are a number of important areas of future study which have not been covered in this work. Firstly, the collection of experimental data on real-world device-to-device encounters is an important area of study for future development of contact tracing protocols. Whilst we used a real-world dataset, there may be differences in human behaviour reflected in a device-to-device contact tracing dataset. Secondly, Bluetooth implementation and battery efficiency for this protocol remains to be studied; for example, techniques may be deployed to increase the frequency of Bluetooth scanning as the duration of a persistent contact increases. Finally, both Bluetooth and cryptographic approaches should be considered to decrease communication overhead as the number of neighbouring devices increase.

\bibliography{references}
\bibliographystyle{ieeetr}
\end{document}